\documentclass[useAMS,usenatbib]{mn2e}

\usepackage{natbib}
\usepackage{graphicx}
\usepackage{multicol}
\usepackage{longtable}
\usepackage{amssymb}
\usepackage[width=1.0\textwidth]{caption}
\usepackage[usenames,dvipsnames]{color}
\RequirePackage[pdftex, plainpages = false, pdfpagelabels]{hyperref}
\definecolor{Red}{rgb}{1.0,0,0}

\newcommand{\etal}    {{\it et al}}                          
\newcommand{\SL}  [2] {$^{#1}$#2}                  
\newcommand{\SLP} [3]{$^{#1}$#2$^{\rm{#3}}$}            
\newcommand{\SLJ} [3]{$^{#1}$#2$_{\rm{#3}}$}            
\newcommand{\AS}    {{\sc Autostructure}}

\newcommand{\II}      {~{\sc ii}}
\newcommand{\III}     {~{\sc iii}}
\newcommand{\IV}     {~{\sc iv}}

\newcommand{\ps} [1]{\overline{#1}}

\title[Atomic data for {\rm Co\III} forbidden lines]
{Collision strengths and transition probabilities for Co\III\ forbidden lines}
\author[P.J. Storey \& T. Sochi]
{P.J. Storey$^{1}$, Taha Sochi$^{1}$\thanks{E-mail: t.sochi@ucl.ac.uk} \\
$^{1}$Department of Physics and Astronomy, University College London, Gower Street, London WC1E
6BT, UK}

\makeatletter \DeclareRobustCommand{\element}[1]{\@element#1\@nil}
\def\@element#1#2\@nil{%
  #1%
  \if\relax#2\relax\else\MakeLowercase{#2}\fi}
 \makeatother

\begin{document}

\date{Accepted 2016 March 31. Received 2016 March 23; in original form 2016 February 01.  \vspace{0.3cm}}

\maketitle

\label{firstpage}

\begin{abstract}
In this paper we compute the collision strengths and their thermally-averaged Maxwellian values for
electron transitions between the fifteen lowest levels of doubly-ionised cobalt, Co$^{2+}$, which
give rise to forbidden emission lines in the visible and infrared region of spectrum. The
calculations also include transition probabilities and predicted relative line emissivities. The
data are particularly useful for analysing the thermodynamic conditions of supernova ejecta.
\vspace{0.3cm}
\end{abstract}

\begin{keywords}
atomic data -- atomic processes -- radiation mechanisms: non-thermal -- supernovae: general --
infrared: general. \vspace{0.4cm}
\end{keywords}

\section{Introduction} \label{Introduction}

Cobalt is an iron-group element but is the least abundant of this group with a solar abundance of
about 300 times less than Fe. However, in supernova (SN) ejecta it is much more abundant. For
example, in SN 1987A the ratio of Co to Fe, 255 days after outburst, is approximately 0.2 by number
\citep{VaraniMSA1990}. The spectral lines of Co are therefore valuable investigative tools in
analysing the chemical and thermodynamic conditions of supernovae where these emissions are mostly
found. These lines are also useful in investigating the evolutionary history and chemical
development by nucleosynthesis and decay processes within the SN explosions \citep{ColgateM1969,
AxelrodThesis1980, KuchnerKPL1994, BowersMGWPe1997, LiuJSQSP1997, ChurazovSIKJe2014,
ChildressHSSMe2015}. The lines of Cobalt have also been observed in the spectral emissions of
astronomical objects with more normal Co abundances such as planetary nebulae
\citep{BaluteauZMP1995, ZhangLLPB2005, PottaschS2005b, WangL2007, FangL2011}.

Little computational and experimental work has been done previously to generate essential atomic
data for Co\III\ and none of the previous work deals with excitation of Co$^{2+}$ levels by
electron impact. \citet{HansenRU1984} calculated magnetic dipole and electric quadrupole transition
probabilities in the 3d$^7$ ground configuration of Co\III\ using parametric fitting to the
observed energy levels and Hartree-Fock values for the electric quadrupole moments. In their
investigation of the forbidden transition probabilities relevant to the analysis of infrared lines
from SN 1987A, \citet{NussbaumerS1988} provided a few transition probabilities for low levels of
Co\III\ assuming $LS$-coupling. \citet{TankosicPD2003} calculated Stark broadening data for a
number of Co\III\ spectral lines as a function of temperature by using a semi-empirical approach.
Experimental investigations have also been conducted by \citet{SugarC1981, SugarC1985} where atomic
data related to Co\III\ transitions, mainly energy levels of Co$^{2+}$, have been collected. Very
recently, \citet{FivetQB2016} calculated radiative probabilities of Co\III\ forbidden transitions
between low-lying levels of doubly ionised cobalt as part of a larger investigation of the
radiative rates in doubly ionised iron-peak elements.

We have recently reported a calculation of atomic parameters for energetically low-lying levels of
Co$^+$ \citep{StoreyZS15}. In this paper we present a similar calculation of atomic parameters
related to forbidden transitions in Co$^{2+}$, which includes lines ranging from the visible to the
three mid-infrared lines which arise from transitions within the ground term at 11.88, 16.39 and
24.06~$\mu$m. The paper primarily addresses a shortage in collisional atomic data which forced some
researchers \citep{DessartHBK2014, ChildressHSSMe2015} to adopt collision strengths generated for
Ni\IV\ \citep{SunderlandNBB2002} as a substitute for corresponding data of Co\III\ justifying this
by the fact that the two ions possess similar electronic and term structures. Our principal result
is collision strengths and their thermally-averaged Maxwellian values for electron excitation and
de-excitation between the fifteen lowest levels of Co$^{2+}$. The study also includes the most
important radiative transition probabilities for the same levels. The main tools used in generating
these data are the R-matrix atomic scattering code \citep{BerringtonBCCRT1974,
BerringtonBBSSTY1987, HummerBEPST1993, BerringtonEN1995}\footnote{{See Badnell: R-matrix write-up
on WWW. URL: \url{amdpp.phys.strath.ac.uk/UK_RmaX/codes/}.}} and the general purpose \AS\ code
\citep{EissnerJN1974, NussbaumerS1978, AS2011}\footnote{{See Badnell: \AS\ write-up on WWW. URL:
\url{amdpp.phys.strath.ac.uk/autos/}.}}. The scattering calculations were performed using a
10-configuration atomic target within a Breit-Pauli intermediate coupling approximation, as will be
detailed in Section~\ref{AtomicStructure}.

The paper is structured as follow. In Section \ref{AtomicStructure} the Co$^{2+}$ model is
presented and the resulting transition probabilities are given, whereas in Section \ref{Scattering}
the Breit-Pauli R-matrix Co$^{2+}+e$ scattering calculation is described. Results and general
analysis related to the diagnostic potentials of some transitions appear in Section \ref{Results},
and section \ref{Conclusions} concludes the paper.

\section{\element{Co}$^{2+}$ Atomic Structure} \label{AtomicStructure}

\subsection{The scattering target}

\begin{figure}
\centering
\includegraphics[height=7cm, width=8.5cm]{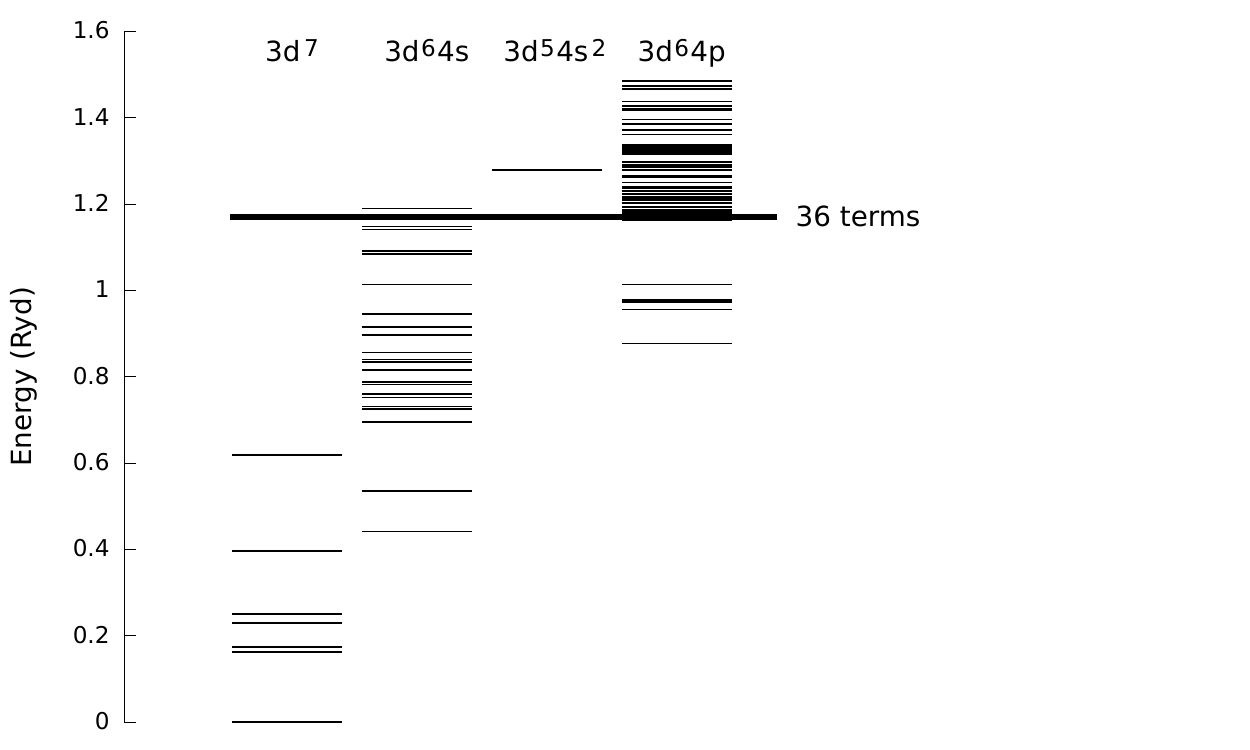}
\caption[]{Schematic term energy diagram of Co\III. The heavy solid line shows the extent of the
close-coupled target states. \label{termdiagram}}
\end{figure}

A schematic diagram of the term structure of Co\III\ up to 1.5~Rydberg is shown in
Figure~\ref{termdiagram}. The extent of our target is shown by the heavy solid line in that figure
and includes 36 terms and 109 levels. The lowest 21 terms of this ion are of even parity from the
configurations 3d$^7$ and 3d$^6$4s. Transitions from higher terms give rise to lines that should be
weaker at the typical temperatures of supernova ejecta and hence they will be ignored. The
odd-parity terms of the 3d$^6$4p configuration are expected to give rise to resonances that affect
the collision strengths for excitation of the low-lying even-parity levels and hence they are
included in the target for the scattering calculations.

A set of ten electron configurations, listed in Table~\ref{configs}, were used to expand the target
states. The target wavefunctions were generated with the \AS\ program, \citep{EissnerJN1974,
NussbaumerS1978, AS2011} using radial functions computed within scaled Thomas-Fermi-Dirac
statistical model potentials. The scaling parameters were determined by minimising the sum of the
energies of all the target terms, computed in $LS$-coupling, i.e. by neglecting all relativistic
effects. The resulting scaling parameters, $\lambda_{nl}$, are given in Table~\ref{scale}.

\begin{table}
\caption{The ten target configuration basis where the core electronic structure ([Ar]) is
suppressed. The bar indicates a correlation orbital.}
\begin{flushleft}
\centering
\begin{tabular}{l}
\noalign{\hrule}
3d$^7$  \\
3d$^6$ 4s, 4p, $\ps4$d  \\
3d$^5$ 4s$^2$, 4p$^2$, $\ps4$d$^2$, 4s4p, 4s$\ps4$d, 4p$\ps4$d  \\
\noalign{\hrule}
\end{tabular}
\end{flushleft}
\label{configs}
\end{table}

\begin{table}
\caption{Potential scaling parameters. The bar over the principal quantum number signifies a
correlation orbital.}
\centering %
\begin{tabular}{lr@{\hskip 1.3cm}lr@{\hskip 1.3cm}lr}
\hline
        1s &    1.42912 &            &            &            &            \\
        2s &    1.13799 &         2p &    1.08143 &            &            \\
        3s &    1.06915 &         3p &    1.05203 &         3d &    1.04962 \\
        4s &    1.03440 &         4p &    1.02977 &     $\ps4$d &   1.51187 \\
\hline
\end{tabular}
\label{scale}
\end{table}

In Table~\ref{termlist} a comparison is made between the term energies calculated using our
scattering target with experimental values for the 36 terms of the target. The term energies are
computed with the inclusion of one-body relativistic effects, the Darwin and mass terms, and the
spin-orbit interaction. This is the type of approximation that we applied for the scattering
calculations in the R-matrix code. In Table~\ref{levellist} the calculated energies of the 15
lowest levels are compared with the corresponding experimental values. The table also shows the
values obtained by including the two-body fine structure interactions as described by
\citet{EissnerJN1974}. The calculated fine-structure splittings of these levels are improved by
this inclusion. For the total fine-structure splitting of the six terms, the average absolute
difference from experiment drops from 7.3\% to 4.6\%.

\begin{table}
\caption{Energies of the 36 target terms in cm$^{-1}$, ordered according to the experimental
energy. The calculated values include only the spin-orbit contribution to the fine-structure
energies. Core electronic structure ([Ar]) is suppressed from all configurations.}
\centering %
\begin{tabular}{llrrl}
\noalign{\hrule}
 & \hspace{1.5cm} & \multicolumn{2}{c}{\hspace{-0.5cm}Term Energy} \cr
 Config. & Term & Exp.$^{\dagger}$ & Calc.  \cr
\noalign{\hrule}
3d$^7$     &     a\SL4F &      0 &          0 \\
           &     a\SL4P &  14561 &      17891 \\
           &     a\SL2G &  16510 &      19120 \\
           &     a\SL2P &  19618 &      25103 \\
           &     a\SL2H &  22227 &      25205 \\
           &     a\SL2D &  22712 &      27507 \\
           &     a\SL2F &  36372 &      43416 \\
3d$^6$4s   &     a\SL6D &  46230 &      48501 \\
           &     a\SL4D &  55448 &      58817 \\
           &     b\SL4P &  70965 &      79599 \\
           &     a\SL4H &  71096 &      76483 \\
           &     b\SL4F &  72717 &      80163 \\
           &     a\SL4G &  76219 &      83370 \\
           &     b\SL2P &  76521 &      85780 \\
           &     b\SL2H &  76690 &      82428 \\
           &     b\SL2F &  78323 &      86408 \\
           &     b\SL2G &  81793 &      89400 \\
           &     b\SL4D &  83031 &      92162 \\
           &     a\SL2I &  84676 &      91484 \\
           &     c\SL2G &  85485 &      93867 \\
           &     b\SL2D &  90897 &      98436 \\
3d$^6$4p   &   z\SLP6Do & 97807  &      97268 \\
3d$^6$4s   &  \phantom{a}\SL2S &        &     100359 \\
3d$^6$4p   &   z\SLP6Fo & 102620 &     102460 \\
3d$^6$4s   &  \phantom{a}\SL2D &        &     103690 \\
3d$^6$4p   &   z\SLP6Po & 104861 &     104906 \\
           &   z\SLP4Do & 106074 &     106802 \\
           &   z\SLP4Fo & 106676 &     107272 \\
           &   z\SLP4Po & 109902 &     111225 \\
3d$^6$4s   &    \phantom{a}\SL2F &        &     111250 \\
           &    \phantom{a}\SL4F &        &     119049 \\
           &    \phantom{a}\SL4P &        &     119600 \\
           &    \phantom{a}\SL2F &        &     125226 \\
           &    \phantom{a}\SL2P &        &     125937 \\
3d$^6$4p   &   z\SLP4So & 122305 &     129103 \\
           &   z\SLP4Go & 124219 &     127494 \\
\noalign{\hrule} \multicolumn{5}{l}{$^{\dagger}$Experimental energies are from NIST (\url{www.nist.gov}).} \\
\end{tabular}
\label{termlist}
\end{table}

\begin{table}
\caption{Energies in cm$^{-1}$ of the 15 lowest levels of Co$^{2+}$, ordered according to the
experimental energy, where the configuration of all levels is [Ar] 3d$^7$.}
 \centering
\begin{tabular}{clrrrr}
\noalign{\hrule}
 Index & Level &
\multicolumn{1}{c}{Exp.$^{1}$} & Calc.$^2$ &  Calc.$^3$ \\
\noalign{\hrule}
       1  & $a^4$F$_{9/2}$     & 0.     & 0.    & 0. \\
       2  & $a^4$F$_{7/2}$     & 841    & 810   & 824 \\
       3  & $a^4$F$_{5/2}$     & 1451   & 1408  & 1428 \\
       4  & $a^4$F$_{3/2}$     & 1867   & 1819  & 1842 \\
       5  & $a^4$P$_{5/2}$     & 15202  & 18481 & 18502 \\
       6  & $a^4$P$_{3/2}$     & 15428  & 18770 & 18785 \\
       7  & $a^4$P$_{1/2}$     & 15811  & 19125 & 19118 \\
       8  & $a^2$G$_{9/2}$     & 16978  & 19565 & 19581 \\
       9  & $a^2$G$_{7/2}$     & 17766  & 20348 & 20357 \\
       10 & $a^2$P$_{3/2}$     & 20195  & 25601 & 25633 \\
       11 & $a^2$P$_{1/2}$     & 20919  & 26486 & 26474 \\
       12 & $a^2$H$_{11/2}$    & 22720  & 25690 & 25687 \\
       13 & $a^2$D$_{5/2}$     & 23059  & 27795 & 27804 \\
       14 & $a^2$H$_{9/2}$     & 23434  & 26367 & 26379 \\
       15 & $a^2$D$_{3/2}$     & 24237  & 29058 & 29033 \\
\noalign{\hrule}
\multicolumn{6}{l}{$^{1}$ \citealt{SugarC1985}.} \\
\multicolumn{6}{l}{$^2$ Calculated with only spin-orbit interaction.} \\
\multicolumn{6}{l}{$^3$ As 2 plus two-body fine-structure interactions} \\
\multicolumn{6}{l}{\,\,\,     for the first 4 configurations of Table~\ref{configs}.} \\
\end{tabular}
\label{levellist}
\end{table}

A widely-accepted measure for the quality of the scattering calculations is the degree of agreement
between weighted oscillator strengths, $gf$, calculated in the velocity and length formulations,
where good agreement is regarded as necessary but not sufficient condition for the quality of the
target wavefunctions. Table~\ref{gflv} provides this comparison where it shows an average
difference in the absolute values of $gf$ of about 5.8\% between the two formulations, which in our
view is acceptable for an open d-shell atomic system.

\begin{table}
\caption{Weighted $LS$ oscillator strengths, $gf$, in the length and velocity formulations from the
two energetically lowest terms of the 3d$^7$ and 3d$^6$4s configurations.}  \centering
\begin{tabular}{llllrr@{\hskip 1.5cm}rr}
\noalign{\hrule}
 & \multicolumn{5}{c}{\hspace{-1.5cm}Transition}  & $gf_{_L}$ & $gf_{_V}$ \\
\noalign{\hrule}
  & 3d$^7$    & $^4$F & -- & 3d$^6$4p & $^4$D$^{\rm o}$ & 2.34  &  2.48      \\
  &           &       & -- &         & $^4$F$^{\rm o}$ & 1.16 & 1.21 \\
  &           &       & -- &         & $^3$G$^{\rm o}$ & 2.38 & 2.30  \\
  & 3d$^6$4s  & $^6$D & -- & 3d$^6$4p & $^6$D$^{\rm o}$ & 9.45 & 9.75 \\
  &           &       & -- &         & $^6$F$^{\rm o}$ & 13.7 & 13.5 \\
  &           &       & -- &         & $^6$P$^{\rm o}$ & 5.77 & 4.83 \\
\noalign{\hrule}
\end{tabular}
\label{gflv}
\end{table}

\subsection{Transition probabilities}\label{Avalues}

The forbidden transition probabilities between the even parity low-lying terms are calculated using
the afore-described target wavefunctions, with empirical adjustments to the computed energies to
ensure more reliable calculation of the fine-structure interactions and accurate energy factors
connecting the {\it ab initio} calculated line strengths to the transition probabilities. The
results for the lowest 15 levels are given in Table~\ref{Avalues2} where the values represent the
sum of the electric quadrupole and magnetic dipole contributions for each transition. This table
includes only those probabilities from a given upper level which exceed 1\% of the total
probability from that level.

The infrared lines of principal interest here arise from transitions between the levels of the
ground \SL4F term and are predominantly of magnetic dipole type. There is therefore a stepwise
decay through the levels and only three relevant transition probabilities, for $a^4$F$_{3/2}$ -
$a^4$F$_{5/2}$, $a^4$F$_{5/2}$ - $a^4$F$_{7/2}$ and $a^4$F$_{7/2}$ - $a^4$F$_{9/2}$. We are aware
of only two previous calculations of transition probabilities for Co\III, one by
\citet{HansenRU1984} and one by \citet{NussbaumerS1988}, as well as one contemporary calculation by
\citet{FivetQB2016}. \citet{NussbaumerS1988} only give values for these three probabilities and
these differ by less than 1\% from our values. \citet{HansenRU1984} give more extensive results
which we compare with the present values in Table~\ref{Avalues2}. We find excellent agreement with
\citet{HansenRU1984} for the magnetic dipole transitions between the levels of individual terms
with differences of a few percent or less. There are larger differences for the electric quadrupole
transition probabilities between terms. For example the probabilities for the principal transitions
between the $a^4$F and $a^4$P terms, the 5-1, 5-2 and 5-3 probabilities, are all larger, by on
average 13\%, in our calculation than in \citet{HansenRU1984}. The fact that all three transitions
differ by approximately the same factor suggests that the cause of the difference lies in the
radial quadrupole integrals used in the two calculations.  There is configuration interaction
between the terms of the 3d$^7$ electron configuration and the 3d$^6\ps4$d configuration in our
calculation and not in the single configuration calculation of \citet{HansenRU1984}. With this
interaction included, the quadrupole line strength involves both the 3d radial quadrupole integral
and the $\ps4$d integral which is significantly larger than for the 3d.

\citet{FivetQB2016} have made calculations of forbidden transition probabilities for the twice
ionised iron-peak elements from Sc to Ni, including Co, and we compare with their results in
Table~\ref{Avalues2}. Their calculations were made with two different methods which we label as
FQB1 and FQB2. The FQB2 values were computed with \AS\ as in the present work. Apart from the
magnetic dipole transitions between the levels of the ground term, which agree to all tabulated
figures, the FQB2 results for the electric quadrupole transitions between levels of different terms
are systematically larger than the present work by 15-20\% with half of them differing by the same
fixed amount of 19\%. As discussed above in the comparison with the work of \citet{HansenRU1984},
the systematic nature of the difference suggests that it is due to a different value for the 3d
radial quadrupole integral rather than details of the wave function expansions of individual terms.
The configuration expansions in the present work and in that of \citet{FivetQB2016} are very
similar but differ in one key aspect. We use a somewhat contracted $\ps4$d orbital to allow for the
differences in the 3d orbital between the 3d$^7$ and 3d$^6$4s configurations, while
\citet{FivetQB2016} employ a spectroscopic 4d orbital but a contracted 5s orbital which provides
flexibility to the spectroscopic 4s. These two different expansions give broadly similar energy
levels and fine-structure but result in differences in the quadrupole radial integrals. It is not
clear that either approach is necessarily superior, so the approximately 15-20\% differences are
probably a realistic measure of the uncertainty in the results for the electric quadrupole line
strengths. We note that the results for the electric quadrupole transition probabilities from the
FQB1 calculation of \citet{FivetQB2016} agree better with their FQB2 for some transitions and
better with the current work for others.

\section{Scattering Calculations} \label{Scattering}

In this work we used the Breit-Pauli R-matrix method, which is detailed in \citet{HummerBEPST1993,
BerringtonEN1995} and the references therein, to perform the scattering calculations. The
calculations were made using the R-matrix codes\footnote{{See Badnell: R-matrix write-up on WWW.
URL: \url{amdpp.phys.strath.ac.uk/UK_RmaX/codes/}.}} where the serial version of the codes were
used in some stages and the parallel version in others. An R-matrix boundary radius of 11.3~au
defining the inner region was applied so that the most extended orbital (4p) of our target is
covered. Each one of the partial waves of the scattered electron was expanded over 12 basis
functions within the R-matrix boundary, and the expansion extends to a maximum of $J=9$.

\begin{figure}
\centering %
{\begin{minipage}[]{0.5\textwidth} \centering \includegraphics[height=3.5cm, width=7.5cm]
{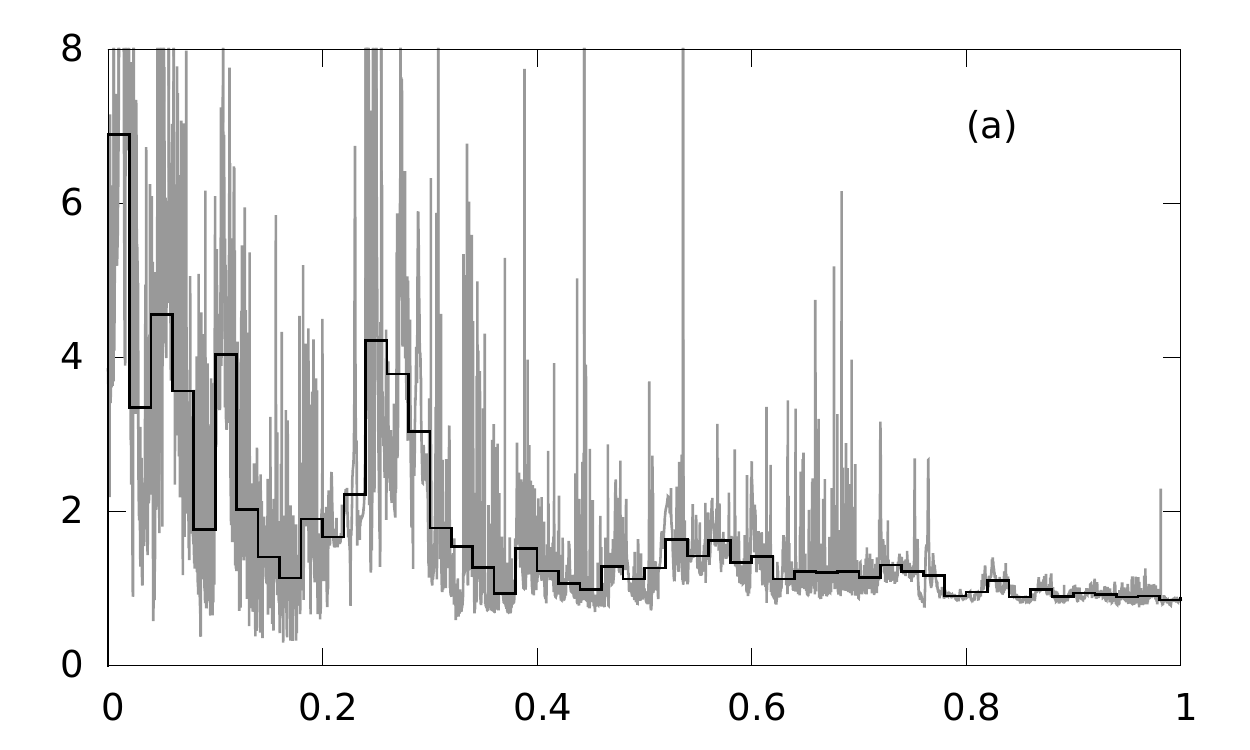}
\end{minipage}} \vspace{0cm}
\centering %
{\begin{minipage}[]{0.5\textwidth} \centering \includegraphics[height=3.5cm, width=7.5cm]
{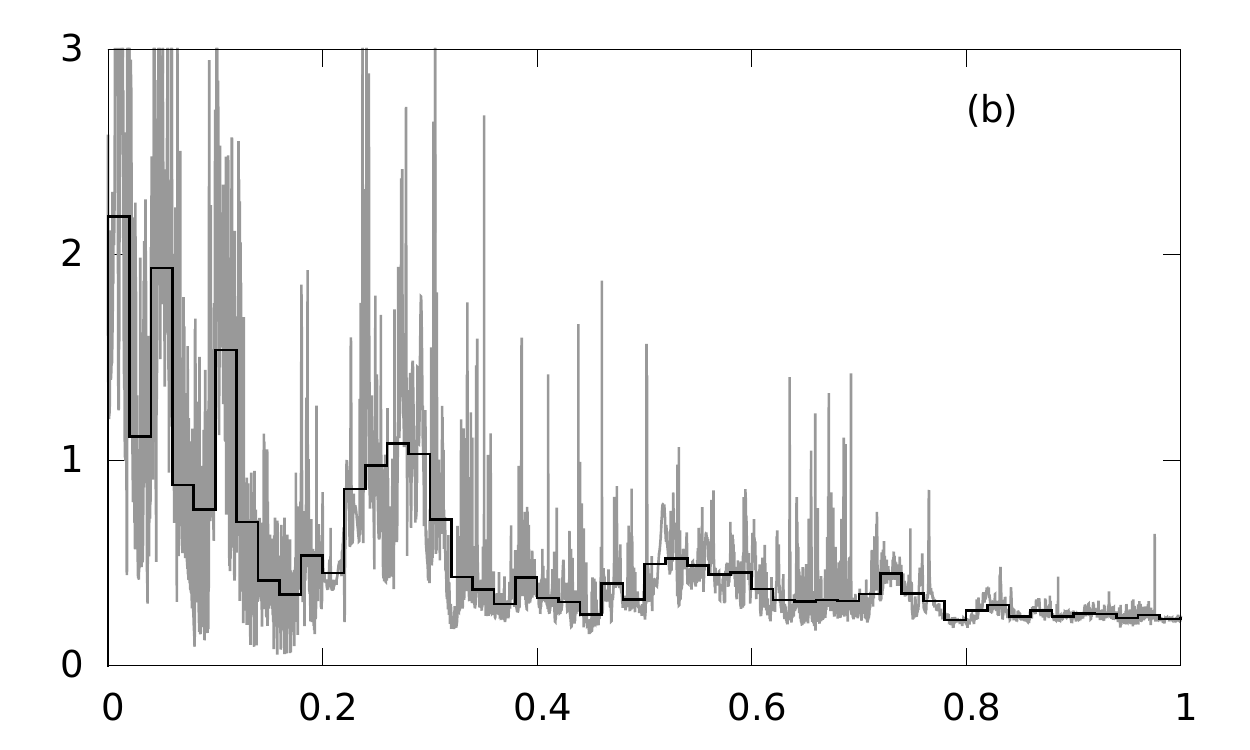}
\end{minipage}} \vspace{0cm}
\centering %
{\begin{minipage}[]{0.5\textwidth} \centering \includegraphics[height=3.5cm, width=7.5cm]
{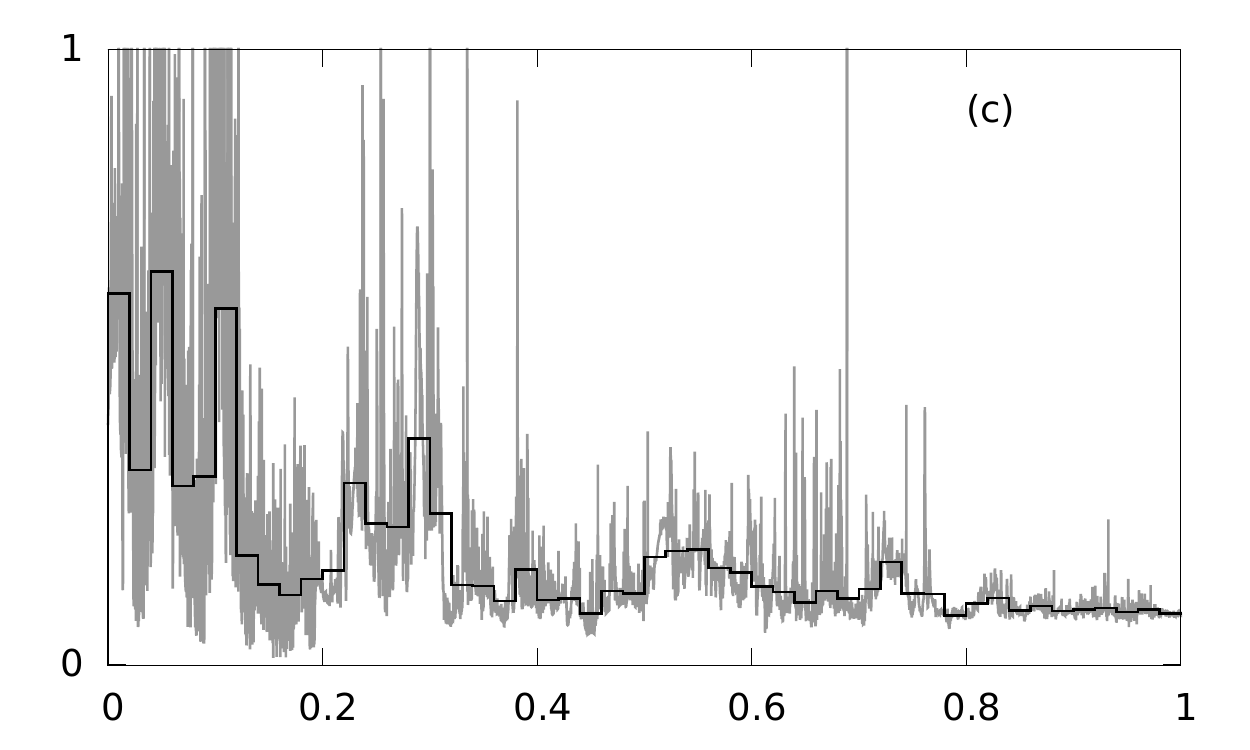}
\end{minipage}}
\centering %
{\begin{minipage}[]{0.5\textwidth} \centering \includegraphics[height=3.5cm, width=7.5cm]
{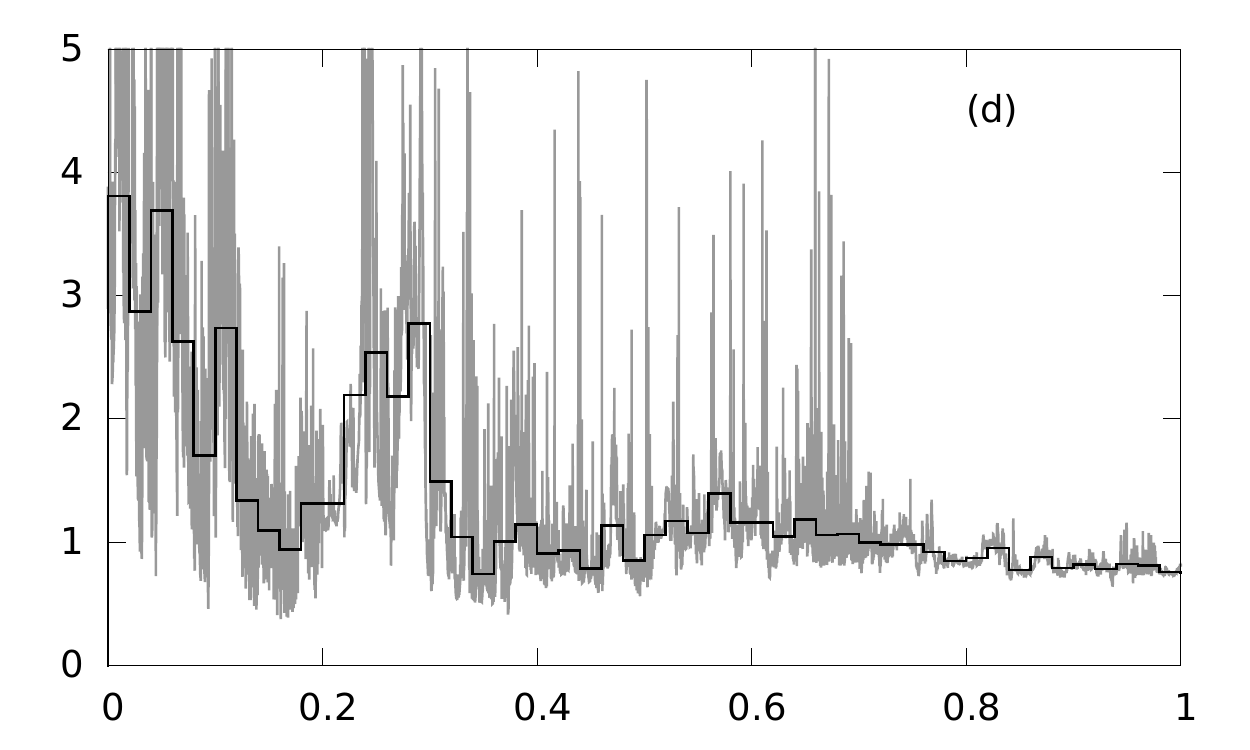}
\end{minipage}}
\centering %
{\begin{minipage}[]{0.5\textwidth} \centering \includegraphics[height=3.5cm, width=7.5cm]
{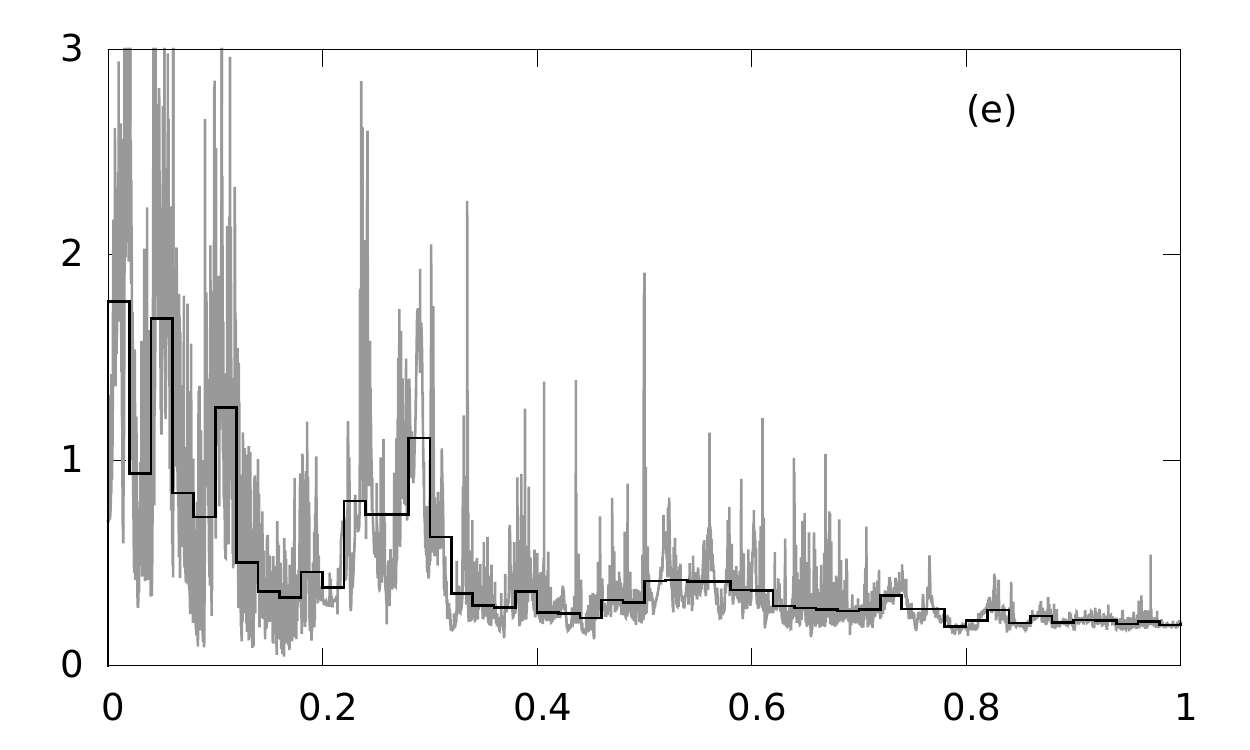}
\end{minipage}}
\centering %
{\begin{minipage}[]{0.5\textwidth} \centering \includegraphics[height=3.5cm, width=7.5cm]
{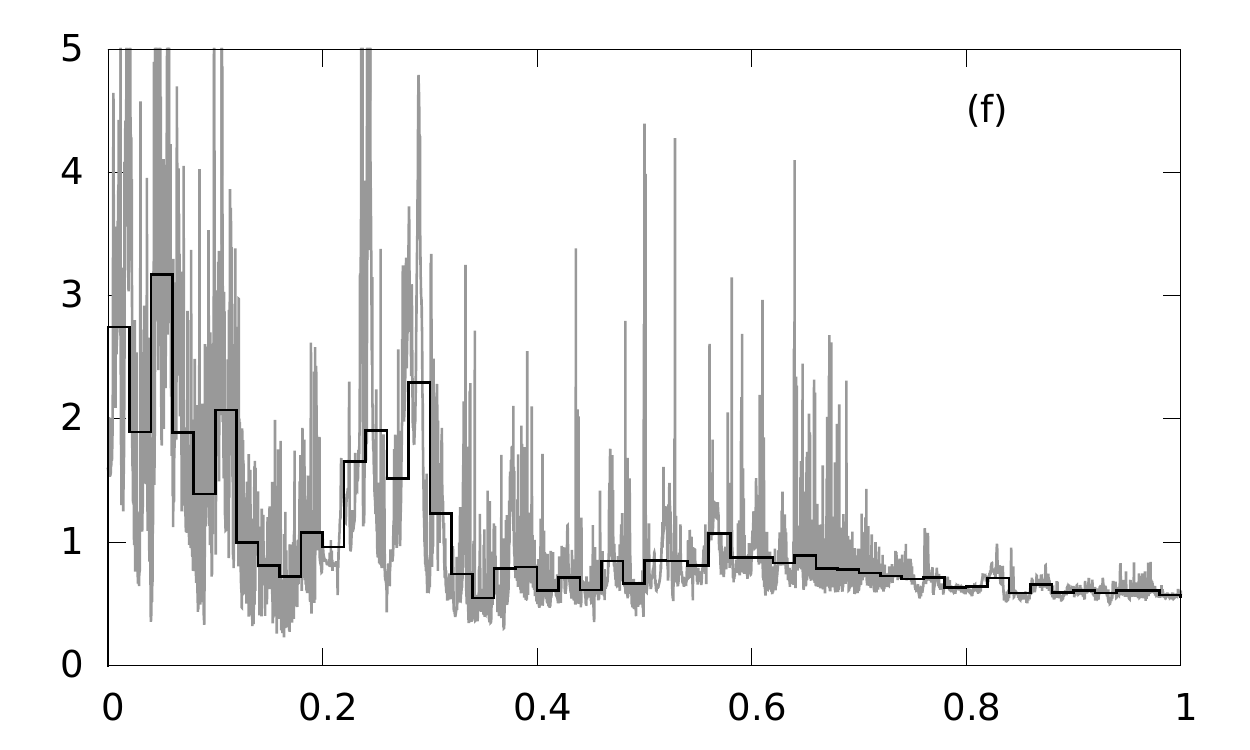}
\end{minipage}}
\caption{Collision strength (vertical axis) versus final electron energy in Rydberg (horizontal
axis) for the (a) 1-2, (b) 1-3, (c) 1-4, (d) 2-3, (e) 2-4 and (f) 3-4 transitions, where the grey
smooth line represents the continuous function while the black discrete line represents binned plot
of the same function to show the magnitude of resonance contributions more clearly. For level
indexing refer to Table~\ref{levellist}. \label{plotOmega}}
\end{figure}

Collision strengths were computed over two non-overlapping energy meshes: a fine mesh consisting of
20000 evenly-divided intervals which goes from zero up to the highest target threshold (about
1.2~Rydberg), and a coarse mesh consisting of 2000 evenly-divided intervals which reach 1~Rydberg
above the highest target threshold. The purpose of the first mesh is to cover the main resonance
region while the second mesh is intended to cover the region where all scattering channels are
open, up to an incident electron energy of about 2.2~Rydberg. Our results demonstrate that these
meshes have achieved these purposes. In Figure~\ref{plotOmega} we illustrate our results with the
computed collision strengths between the lowest four levels of the ground 3d$^7$~$^4$F term as a
function of final electron energy up to 1~Rydberg above threshold. Dense and complex resonance
structure can be seen in these plots due to the multiple close lying thresholds. We also show the
collision strength averaged over 0.02~Rydberg intervals.

To ensure that the computed collision strengths have converged in partial wave for all the levels
for which data are given, the contribution of partial wave $J=9$ was compared to the sum for all
transitions and energies. This comparison showed that in almost all cases the contribution from
$J=9$ is negligible. Specifically, the largest contribution from $J=9$ is for the transition 8-12
at about 1\% and the next largest is about 0.1\% of the total. However, we note that it is certain
that the collision strengths from the lower levels to the levels of the 4p configuration are not
converged because they are allowed transitions which have significant high partial wave
contributions. We therefore do not provide collision strengths for any of these transitions.

\section{Results and Discussion} \label{Results}

The thermally-averaged collision strengths between the fifteen lowest energy levels are given in
Table~\ref{upstable} as a function of electron temperature. These values were calculated using the
full energy range, as described above. In the energy region where all scattering channels are open
there are some small irregular features in the collision strengths that are almost certainly
non-physical and caused by the correlation orbital in the target representation. We computed
thermally-averaged collision strengths for the transitions and temperature range given in
Table~\ref{upstable} both including and excluding the contribution from the region of all channels
open, and found the largest change for any transition is 0.3\% at log$_{10}T=4.0$, 2.4\% at
log$_{10}T=4.2$ and 9.4\% at log$_{10}T=4.4$. The values tabulated in Table~\ref{upstable} were
computed using the full energy range.

\subsection{Principal spectral lines}\label{irratios}

We compute the predicted Co$^{2+}$ fractional level populations using the results in
Tables~\ref{Avalues2} and \ref{upstable} with a fifteen level model atom including electron
collisional excitation and de-excitation and radiative decay. In Tables~\ref{coiiilinesPN} and
\ref{coiiilinesSN} we show the resulting ten strongest lines of Co\III\ in this model. We also
ensure that the three Co\III\ mid-infrared lines at $11.88, 16.39$ and $24.06$~$\mu$m are in the
tables even if they are not among the ten strongest. The fifteen levels are all of even parity so
all these lines are [Co\III] forbidden transitions. The tabulated quantity $\rho$ is the ratio of
the energy emitted per unit time in a Co\III\ line relative to H$\beta$ for unit Co$^{2+}$ and
H$^+$ ion number density. Hence for a downward transition of wavelength $\lambda_{ij}$ between
Co$^{2+}$ levels $j$ and $i$,
\begin{equation}
\rho = \frac{f_j A_{ji} \lambda_{{\rm H}\beta}}{N_e \alpha_{e}^{{\rm H}\beta} \lambda_{ij}}
\end{equation}
where $f_j$ is the fraction of Co$^{2+}$ in the upper state $j$, $A_{ji}$ is the Einstein
A-coefficient for the transition, $\lambda_{{\rm H}\beta}$ is the H$\beta$ wavelength and
$\alpha_{e}^{{\rm H}\beta}$ is the effective recombination coefficient for H$\beta$ whose value is
obtained from \citet{StoreyH95}. The values of $\rho$ are tabulated for a temperature of $10^4$~K
and for two electron densities, $N_e=10^4$~cm$^{-3}$ typical of planetary nebulae
(Table~\ref{coiiilinesPN}), and $N_e=10^7$~cm$^{-3}$ more typical of SN remnants in their nebular
phase (Table~\ref{coiiilinesSN}). Thus, in typical PN conditions, assuming a Co abundance of
$10^{-7}$ with respect to H$^+$ by number and assuming 20\% of Co is in the form of Co$^{2+}$, the
brightest visible Co\III\ line at $5888.5$~\AA\ would have an emissivity per unit volume $5.7\times
10^{-4}$ times that of H$\beta$. In principle this would be visible in deep spectra of bright PNe
(e.g. \citet{BaluteauZMP1995}). In practice, \citet{BaluteauZMP1995} do not identify this line in
the spectrum of NGC~7027 which may reflect depletion of gas phase Co on dust grains.

In \cite{StoreyZS15} we reported collision strengths and transition probabilities for low-lying
transitions in Co\II\ and discussed the spectroscopic uses of the three mid-infrared lines at
$10.52, 14.74$ and $15.46$~$\mu$m. There are also significant Co\II\ visible and near-infrared
lines which were not discussed by \cite{StoreyZS15}, so in Tables~\ref{coiilinesPN} and
\ref{coiilinesSN} we show the strongest of these. The Co\II\ model atom also comprises the
energetically lowest 15 levels and the transition probabilities and thermally-averaged collision
strengths required are all from \cite{StoreyZS15}.

\begin{table}
\caption{The emissivity ratio, $\rho$, of the 10 strongest lines of Co\III\ in our 15-level model
atom for electron temperature $T_e=10^4$~K and electron number density $N_e=10^4$~cm$^{-3}$ typical
of PNe. We also add the $24.06~\mu$m line to show the relative strength of all three mid-infrared
lines. The powers of 10 of the $\rho$ values are given in brackets and $i$ and $j$ refer to the
lower and upper levels respectively as indexed in Table \ref{levellist}.}
 \centering
\begin{tabular}{rrlll@{\hskip 0.5cm}rrrr}
\noalign{\hrule}
 $j$ & $i$ & \multicolumn{3}{c}{\hspace{-0.5cm}Transition}  & $\lambda$ \hspace{0.4cm} & $\rho$ \hspace{0.4cm} \\
\noalign{\hrule}
 2 & 1 & 3d$^7$~\SLJ4F{7/2}  &  -- & 3d$^7$~\SLJ4F{9/2} & 11.88$\mu$m & 5.85(+4)  &       \\
 8 & 1 &  3d$^7$~\SLJ2G{9/2}  &  -- & 3d$^7$~\SLJ4F{9/2} & 5888.48\AA & 2.83(+4)  &       \\
 5 & 1 &  3d$^7$~\SLJ4P{5/2}  &  -- & 3d$^7$~\SLJ4F{9/2} & 6576.31\AA & 2.27(+4)  &       \\
 3 & 2 &  3d$^7$~\SLJ4F{5/2}  &  -- & 3d$^7$~\SLJ4F{7/2} & 16.39$\mu$m & 1.26(+4)  &       \\
 8 & 2 &  3d$^7$~\SLJ2G{9/2}  &  -- & 3d$^7$~\SLJ4F{7/2} & 6195.45\AA & 8.47(+3)  &       \\
 6 & 2 &  3d$^7$~\SLJ4P{3/2}  &  -- & 3d$^7$~\SLJ4F{7/2} & 6853.53\AA & 6.93(+3)  &       \\
 5 & 2 &  3d$^7$~\SLJ4P{5/2}  &  -- & 3d$^7$~\SLJ4F{7/2} & 6961.53\AA & 5.83(+3)  &       \\
 13 & 2 &  3d$^7$~\SLJ2D{5/2}  &  -- & 3d$^7$~\SLJ4F{7/2} & 4499.67\AA & 4.07(+3)  &       \\
 6 & 3 &  3d$^7$~\SLJ4P{3/2}  &  -- & 3d$^7$~\SLJ4F{5/2} & 7152.69\AA & 3.91(+3)  &       \\
 12 & 8 &  3d$^7$~\SLJ2H{11/2}  &  -- & 3d$^7$~\SLJ2G{9/2} & 1.741$\mu$m & 3.82(+3)  &       \\
 4 & 3 &  3d$^7$~\SLJ4F{3/2}  &  -- & 3d$^7$~\SLJ4F{5/2} & 24.06$\mu$m & 1.94(+3)  &       \\
\noalign{\hrule}
\end{tabular}
\label{coiiilinesPN}
\end{table}

\begin{table}
\caption{The emissivity ratio, $\rho$, of the 10 strongest lines plus the three mid-infrared lines
of Co\III\ with $N_e=10^7$~cm$^{-3}$ typical of SN remnants. The other details are as in Table
\ref{coiiilinesPN}.}
 \centering
\begin{tabular}{rrlll@{\hskip 0.5cm}rrrr}
\noalign{\hrule}
 $j$ & $i$ & \multicolumn{3}{c}{\hspace{-0.5cm}Transition}  & $\lambda$ \hspace{0.4cm} & $\rho$ \hspace{0.4cm} \\
\noalign{\hrule}
 8 & 1 &  3d$^7$~\SLJ2G{9/2}  &  -- & 3d$^7$~\SLJ4F{9/2} & 5888.48\AA & 1.26(+4)  &       \\
 13 & 2 &  3d$^7$~\SLJ2D{5/2}  &  -- & 3d$^7$~\SLJ4F{7/2} & 4499.67\AA & 5.39(+3)  &       \\
 9 & 2 &  3d$^7$~\SLJ2G{7/2}  &  -- & 3d$^7$~\SLJ4F{7/2} & 5906.78\AA & 3.82(+3)  &       \\
 8 & 2 &  3d$^7$~\SLJ2G{9/2}  &  -- & 3d$^7$~\SLJ4F{7/2} & 6195.45\AA & 3.78(+3)  &       \\
 9 & 3 &  3d$^7$~\SLJ2G{7/2}  &  -- & 3d$^7$~\SLJ4F{5/2} & 6127.67\AA & 2.74(+3)  &       \\
 5 & 1 &  3d$^7$~\SLJ4P{5/2}  &  -- & 3d$^7$~\SLJ4F{9/2} & 6576.31\AA & 2.62(+3)  &       \\
 15 & 3 &  3d$^7$~\SLJ2D{3/2}  &  -- & 3d$^7$~\SLJ4F{5/2} & 4387.52\AA & 2.38(+3)  &       \\
 15 & 4 &  3d$^7$~\SLJ2D{3/2}  &  -- & 3d$^7$~\SLJ4F{3/2} & 4469.02\AA & 1.24(+3)  &       \\
 6 & 2 &  3d$^7$~\SLJ4P{3/2}  &  -- & 3d$^7$~\SLJ4F{7/2} & 6853.53\AA & 9.23(+2)  &       \\
 14 & 8 &  3d$^7$~\SLJ2H{9/2}  &  -- & 3d$^7$~\SLJ2G{9/2} & 1.548$\mu$m & 7.49(+2)  &       \\
 2 & 1 & 3d$^7$~\SLJ4F{7/2}  &  -- & 3d$^7$~\SLJ4F{9/2} & 11.88$\mu$m & 6.76(+2)  &       \\
 3 & 2 &  3d$^7$~\SLJ4F{5/2}  &  -- & 3d$^7$~\SLJ4F{7/2} & 16.39$\mu$m & 2.17(+2)  &       \\
 4 & 3 &  3d$^7$~\SLJ4F{3/2}  &  -- & 3d$^7$~\SLJ4F{5/2} & 24.06$\mu$m & 3.26(+1)  &       \\
\noalign{\hrule}
\end{tabular}
\label{coiiilinesSN}
\end{table}

\begin{table}
\caption{The emissivity ratio, $\rho$, of the 10 strongest lines of Co\II\ with
$N_e=10^4$~cm$^{-3}$ typical of PNe. We also add the $15.46~\mu$m line discussed by
\citet{StoreyZS15}. The other details are as in Table \ref{coiiilinesPN}.}
 \centering
\begin{tabular}{rrlll@{\hskip 0.5cm}rrrr}
\noalign{\hrule}
 $j$ & $i$ & \multicolumn{3}{c}{\hspace{-0.5cm}Transition}  & $\lambda$ \hspace{0.4cm} & $\rho$ \hspace{0.4cm} \\
\noalign{\hrule}
 9 & 1 & 3d$^7$4s~\SLJ3F{4}  &  -- & 3d$^8$~\SLJ3F{4} & 1.019$\mu$m & 1.17(+5)  &       \\
 9 & 4 &  3d$^7$4s~\SLJ3F{4}  &  -- & 3d$^7$4s~\SLJ5F{5} & 1.547$\mu$m & 6.53(+4)  &       \\
 2 & 1 &  3d$^8$~\SLJ3F{3}  &  -- & 3d$^8$~\SLJ3F{4} & 10.52$\mu$m & 3.81(+4)  &       \\
 5 & 4 &  3d$^7$4s~\SLJ5F{4}  &  -- & 3d$^7$4s~\SLJ5F{5} & 14.74$\mu$m & 2.86(+4)  &       \\
 12 & 2 &  3d$^8$~\SLJ1D{2}  &  -- & 3d$^8$~\SLJ3F{3} & 9342.56\AA & 1.74(+4)  &       \\
 9 & 2 &  3d$^7$4s~\SLJ3F{4}  &  -- & 3d$^8$~\SLJ3F{3} & 1.128$\mu$m & 1.27(+4)  &       \\
 9 & 6 &  3d$^7$4s~\SLJ3F{4}  &  -- & 3d$^7$4s~\SLJ5F{3} & 1.903$\mu$m & 9.33(+3)  &       \\
 13 & 2 &  3d$^8$~\SLJ3P{2}  &  -- & 3d$^8$~\SLJ3F{3} & 8121.13\AA & 9.04(+3)  &       \\
 12 & 3 &  3d$^8$~\SLJ1D{2}  &  -- & 3d$^8$~\SLJ3F{2} & 9943.60\AA & 8.13(+3)  &       \\
 10 & 2 &  3d$^7$4s~\SLJ3F{3}  &  -- & 3d$^8$~\SLJ3F{3} & 1.025$\mu$m & 7.27(+3)  &       \\
 3 & 2 &  3d$^8$~\SLJ3F{2}  &  -- & 3d$^8$~\SLJ3F{3} & 15.46$\mu$m & 5.00(+3)  &       \\
\noalign{\hrule}
\end{tabular}
\label{coiilinesPN}
\end{table}

\begin{table}
\caption{The emissivity ratio, $\rho$, of the 10 strongest lines of Co\II\ with
$N_e=10^7$~cm$^{-3}$ typical of SN remnants. We also add the $10.52, 14.74$ and $15.46~\mu$m lines
discussed by \citet{StoreyZS15}. The other details are as in Table \ref{coiiilinesPN}.}
 \centering
\begin{tabular}{rrlll@{\hskip 0.5cm}rrrr}
\noalign{\hrule}
 $j$ & $i$ & \multicolumn{3}{c}{\hspace{-0.5cm}Transition}  & $\lambda$ \hspace{0.4cm} & $\rho$ \hspace{0.4cm} \\
\noalign{\hrule}
 12 & 2 &  3d$^8$~\SLJ1D{2}  &  -- & 3d$^8$~\SLJ3F{3} & 9342.56\AA & 4.25(+3)  &       \\
 9 & 1 & 3d$^7$4s~\SLJ3F{4}  &  -- & 3d$^8$~\SLJ3F{4} & 1.019$\mu$m & 2.41(+3)  &       \\
 12 & 3 &  3d$^8$~\SLJ1D{2}  &  -- & 3d$^8$~\SLJ3F{2} & 9943.60\AA & 1.99(+3)  &       \\
 13 & 2 &  3d$^8$~\SLJ3P{2}  &  -- & 3d$^8$~\SLJ3F{3} & 8121.13\AA & 1.72(+3)  &       \\
 9 & 4 &  3d$^7$4s~\SLJ3F{4}  &  -- & 3d$^7$4s~\SLJ5F{5} & 1.547$\mu$m & 1.35(+3)  &       \\
 10 & 2 &  3d$^7$4s~\SLJ3F{3}  &  -- & 3d$^8$~\SLJ3F{3} & 1.025$\mu$m & 1.07(+3)  &       \\
 13 & 1 &  3d$^8$~\SLJ3P{2}  &  -- & 3d$^8$~\SLJ3F{4} & 7539.01\AA & 9.27(+2)  &       \\
 11 & 2 &  3d$^7$4s~\SLJ3F{2}  &  -- & 3d$^8$~\SLJ3F{3} & 9639.21\AA & 8.72(+2)  &       \\
 11 & 3 &  3d$^7$4s~\SLJ3F{2}  &  -- & 3d$^8$~\SLJ3F{2} & 1.028$\mu$m & 8.70(+2)  &       \\
 10 & 1 &  3d$^7$4s~\SLJ3F{3}  &  -- & 3d$^8$~\SLJ3F{4} & 9335.84\AA & 7.67(+2)  &       \\
 2 & 1 &  3d$^8$~\SLJ3F{3}  &  -- & 3d$^8$~\SLJ3F{4} & 10.52$\mu$m & 4.46(+2)  &       \\
 5 & 4 &  3d$^7$4s~\SLJ5F{4}  &  -- & 3d$^7$4s~\SLJ5F{5} & 14.74$\mu$m & 1.41(+2)  &       \\
 3 & 2 &  3d$^8$~\SLJ3F{2}  &  -- & 3d$^8$~\SLJ3F{3} & 15.46$\mu$m & 8.48(+1)  &       \\
\noalign{\hrule}
\end{tabular}
\label{coiilinesSN}
\end{table}

\section{Conclusions} \label{Conclusions}

In this study, the Co\III\ forbidden lines arising from transitions between the fifteen lowest
energy levels of doubly-ionised cobalt, Co$^{2+}$, have been investigated. Radiative transition
probabilities and collision strengths for excitation and de-excitation by electron scattering, with
their thermally-averaged values based on a Maxwell-Boltzmann statistics, have been computed and
reported. The scattering calculations used the R-matrix method in the Breit-Pauli approximation
under an intermediate coupling scheme.

The emissivities of the Co\III\ forbidden lines were calculated with a 15-level Co$^{2+}$ model
atom and the strongest lines listed with their expected strength relative to H$\beta$ for
conditions approximately representative of those in planetary nebulae and supernova remnants. For
comparison and completeness we also listed the strongest forbidden lines from Co\II\ in the same
conditions based on atomic parameters calculated and presented in a previous paper
\citep{StoreyZS15}.

\section{Acknowledgments \& Statement}

The authors would like to thank the reviewer for drawing their attention to the recent work of
\citet{FivetQB2016} and suggesting the comparison. PJS acknowledges financial support from the
Atomic Physics for Astrophysics Project (APAP) funded by the Science and Technology Facilities
Council (STFC). The collision strength data, as a function of electron energy, for the lowest 15
levels of Co$^{+}$ and Co$^{2+}$ can be obtained in electronic format with full precision from the
Centre de Donn\'{e}es astronomiques de Strasbourg (CDS) database.

\onecolumn

\begin{table}
\caption{Transition probabilities in s$^{-1}$ among the energetically lowest 15 levels of Co$^{2+}$
as obtained from the current work (CW), from \citet{HansenRU1984} (HRU), from \citet{FivetQB2016}
using HFR (FQB1) and from \citet{FivetQB2016} using \AS\ (FQB2). The transition indices $i$ and
$j$, which refer to the lower and upper levels respectively, are as in Table~\ref{levellist}. Only
the CW transition probabilities that are at least 1\% of the total probability from a given upper
level are listed. The powers of 10 by which the numbers are to be multiplied are given in
brackets.}
 \centering
\begin{tabular}{ccccccccccccc}
\hline
\multicolumn{2}{c}{Transition} & \multicolumn{ 4}{c}{A-value} & & \multicolumn{2}{c}{Transition} & \multicolumn{ 4}{c}{A-value} \\

         $j$ &          $i$ &         CW &        HRU & FQB1 & FQB2 &  &       $j$ &          $i$ &         CW &        HRU & FQB1 & FQB2  \\
\hline
2  &          1 &   2.00(-2) & 2.0(-2)  & 2.01(-2) & 2.00(-2) &            &         11 &          3 &   2.23(-3) & 2.2(-3) &          &          \\
3  &          2 &   1.31(-2) & 1.3(-2)  & 1.32(-2) & 1.31(-2) &            &         11 &          4 &   2.69(-3) & 2.4(-3) &          &          \\
4  &          3 &   4.63(-3) & 4.7(-3)  & 4.65(-3) & 4.63(-3) &            &         11 &          7 &   1.77(-1) & 2.0(-1) & 1.98(-1) & 2.01(-1) \\
5  &          1 &   5.55(-2) & 4.8(-2)  & 6.59(-2) & 6.65(-2) &            &         11 &         10 &   6.42(-3) & 6.4(-3) &          &          \\
5  &          2 &   1.51(-2) & 1.35(-2) & 1.74(-2) & 1.78(-2) &            &         12 &          1 &   6.02(-4) & 6.2(-4) &          &          \\
5  &          3 &   3.14(-3) & 2.68(-3) &          &          &            &         12 &          8 &   3.94(-2) & 4.2(-2) & 4.29(-2) & 4.69(-2) \\
6  &          2 &   3.14(-2) & 2.7(-2)  & 3.69(-2) & 3.73(-2) &            &         13 &          2 &   7.34(-1) & 7.5(-1) & 7.44(-1) & 8.27(-1) \\
6  &          3 &   1.85(-2) & 1.63(-2) & 2.18(-2) & 2.21(-2) &            &         13 &          3 &   7.94(-2) & 8.1(-2) &          &          \\
6  &          4 &   5.14(-3) & 4.41(-3) &          &          &            &         13 &          4 &   3.65(-2) & 3.5(-2) &          &          \\
7  &          3 &   2.30(-2) & 2.0(-2)  & 2.71(-2) & 2.73(-2) &            &         13 &          5 &   4.74(-2) & 4.7(-2) &          &          \\
7  &          4 &   3.02(-2) & 2.6(-2)  & 3.57(-2) & 3.60(-2) &            &         13 &          6 &   2.38(-2) & 2.4(-2) &          &          \\
7  &          6 &   2.45(-3) & 2.5(-3)  &          &          &            &         13 &         10 &   1.87(-2) & 1.8(-2) &          &          \\
8  &          1 &   3.71(-1) & 4.0(-1)  & 3.91(-1) & 4.34(-1) &            &         14 &          1 &   3.61(-3) & 4.32(-3)&          &          \\
8  &          2 &   1.17(-1) & 1.2(-1)  & 1.22(-1) & 1.36(-1) &            &         14 &          2 &   1.90(-3) & 2.24(-3)&          &          \\
9  &          1 &   1.38(-2) & 1.4(-2)  &          &          &            &         14 &          8 &   1.23(-1) & 1.3(-1) & 1.33(-1) & 1.46(-1) \\
9  &          2 &   1.40(-1) & 1.5(-1)  & 1.50(-1) & 1.67(-1) &            &         14 &          9 &   3.70(-2) & 3.9(-2) & 4.03(-2) & 4.41(-2) \\
9  &          3 &   1.04(-1) & 1.1(-1)  & 1.12(-1) & 1.24(-1) &            &         14 &         12 &   5.26(-3) & 5.3(-3) &          &          \\
9  &          8 &   7.19(-3) & 7.2(-3)  &          &          &            &         15 &          3 &   6.93(-1) & 7.3(-1) & 7.34(-1) & 8.02(-1) \\
10 &          2 &   5.36(-3) & 5.1(-3)  &          &          &            &         15 &          4 &   3.67(-1) & 3.9(-1) & 3.86(-1) & 4.19(-1) \\
10 &          3 &   6.52(-2) & 6.43(-2) & 6.20(-2) & 8.08(-2) &            &         15 &          6 &   1.52(-2) & 1.4(-2) &          &          \\
10 &          4 &   4.64(-2) & 4.46(-2) & 4.27(-2) & 5.52(-2) &            &         15 &         10 &   1.49(-1) & 1.5(-1) & 1.41(-1) & 1.67(-1) \\
10 &          5 &   1.41(-1) & 1.5(-1)  & 1.55(-1) & 1.58(-1) &            &         15 &         11 &   2.71(-2) & 2.7(-2) &          &          \\
10 &          6 &   7.26(-2) & 8.0(-2)  & 8.04(-2) & 8.08(-2) &            &         15 &         13 &   2.43(-2) & 2.5(-2) &          &          \\
10 &          7 &   3.01(-2) & 3.3(-2)  &          &          &            &            &            &            &         &     &     \\
\hline
\end{tabular}
\label{Avalues2}
\end{table}


\clearpage

\begin{longtable}{cccccccccccccccc}
\caption{Thermally-averaged collision strengths among the 15 energetically lowest levels of
Co$^{2+}$ as a function of log$_{10}$ of temperature in Kelvin where $i$ and $j$ refer to the index
of the lower and upper level respectively (see
Table~\ref{levellist} for indexing). \label{upstable}} \vspace{0cm}\\
\hline
        & $i$  &      $j$      &                                                                                                                                        \multicolumn{ 13}{c}{log$_{10}T$} \\
\cline{4-16}
       &   &           &          2.0 &        2.2 &        2.4 &        2.6 &        2.8 &          3.0 &        3.2 &        3.4 &        3.6 &        3.8 &          4.0 &        4.2 &        4.4 \\
\hline
\endfirsthead
\caption[]{continued.}\\
\hline
         & $i$ &       $j$     &                                                                                                                                        \multicolumn{ 13}{c}{log$_{10}T$} \\
\cline{4-16}
        &  &           &          2.0 &        2.2 &        2.4 &        2.6 &        2.8 &          3.0 &        3.2 &        3.4 &        3.6 &        3.8 &          4.0 &        4.2 &        4.4 \\
\hline
\endhead
\hline
\endfoot
 &  1 &  2 & 4.037 & 4.171 & 4.321 & 4.573 & 5.001 & 5.470 & 5.699 & 5.586 & 5.229 & 4.732 & 4.177 & 3.636 & 3.135 \\
 &  1 &  3 & 1.490 & 1.471 & 1.511 & 1.630 & 1.795 & 1.926 & 1.957 & 1.893 & 1.769 & 1.607 & 1.419 & 1.228 & 1.045 \\
 &  1 &  4 & 0.429 & 0.448 & 0.473 & 0.502 & 0.528 & 0.545 & 0.544 & 0.529 & 0.505 & 0.473 & 0.429 & 0.378 & 0.325 \\
 &  1 &  5 & 1.285 & 1.328 & 1.379 & 1.409 & 1.404 & 1.364 & 1.299 & 1.226 & 1.177 & 1.182 & 1.221 & 1.243 & 1.224 \\
 &  1 &  6 & 0.578 & 0.611 & 0.626 & 0.618 & 0.590 & 0.551 & 0.509 & 0.471 & 0.456 & 0.472 & 0.497 & 0.506 & 0.490 \\
 &  1 &  7 & 0.232 & 0.215 & 0.201 & 0.188 & 0.178 & 0.171 & 0.164 & 0.156 & 0.155 & 0.160 & 0.167 & 0.168 & 0.162 \\
 &  1 &  8 & 0.963 & 0.944 & 0.937 & 0.949 & 0.973 & 0.997 & 1.027 & 1.071 & 1.115 & 1.154 & 1.193 & 1.225 & 1.235 \\
 &  1 &  9 & 0.312 & 0.322 & 0.315 & 0.299 & 0.282 & 0.270 & 0.267 & 0.274 & 0.283 & 0.292 & 0.302 & 0.309 & 0.308 \\
 &  1 & 10 & 0.361 & 0.377 & 0.404 & 0.422 & 0.421 & 0.407 & 0.393 & 0.384 & 0.373 & 0.361 & 0.349 & 0.340 & 0.332 \\
 &  1 & 11 & 0.180 & 0.166 & 0.147 & 0.127 & 0.109 & 0.094 & 0.084 & 0.077 & 0.073 & 0.072 & 0.073 & 0.077 & 0.081 \\
 &  1 & 12 & 4.532 & 3.989 & 3.412 & 2.884 & 2.444 & 2.093 & 1.805 & 1.570 & 1.407 & 1.330 & 1.316 & 1.327 & 1.328 \\
 &  1 & 13 & 0.375 & 0.369 & 0.374 & 0.393 & 0.417 & 0.430 & 0.431 & 0.428 & 0.431 & 0.442 & 0.463 & 0.486 & 0.500 \\
 &  1 & 14 & 0.374 & 0.374 & 0.364 & 0.342 & 0.313 & 0.284 & 0.259 & 0.242 & 0.238 & 0.244 & 0.254 & 0.261 & 0.260 \\
 &  1 & 15 & 0.070 & 0.066 & 0.065 & 0.066 & 0.068 & 0.070 & 0.073 & 0.075 & 0.078 & 0.081 & 0.086 & 0.090 & 0.092 \\
 &  2 &  3 & 3.301 & 3.280 & 3.245 & 3.264 & 3.382 & 3.536 & 3.617 & 3.581 & 3.442 & 3.209 & 2.905 & 2.578 & 2.258 \\
 &  2 &  4 & 0.732 & 0.760 & 0.831 & 0.962 & 1.139 & 1.315 & 1.428 & 1.457 & 1.415 & 1.319 & 1.184 & 1.034 & 0.884 \\
 &  2 &  5 & 1.089 & 1.064 & 1.046 & 1.026 & 0.987 & 0.930 & 0.864 & 0.804 & 0.772 & 0.785 & 0.820 & 0.838 & 0.816 \\
 &  2 &  6 & 0.658 & 0.682 & 0.695 & 0.690 & 0.668 & 0.639 & 0.606 & 0.571 & 0.544 & 0.540 & 0.552 & 0.559 & 0.551 \\
 &  2 &  7 & 0.292 & 0.274 & 0.254 & 0.233 & 0.215 & 0.203 & 0.192 & 0.183 & 0.185 & 0.202 & 0.223 & 0.232 & 0.227 \\
 &  2 &  8 & 0.617 & 0.610 & 0.602 & 0.601 & 0.603 & 0.605 & 0.609 & 0.625 & 0.648 & 0.673 & 0.698 & 0.716 & 0.719 \\
 &  2 &  9 & 0.490 & 0.487 & 0.474 & 0.461 & 0.451 & 0.447 & 0.454 & 0.472 & 0.489 & 0.501 & 0.512 & 0.521 & 0.524 \\
 &  2 & 10 & 0.231 & 0.230 & 0.240 & 0.258 & 0.275 & 0.284 & 0.284 & 0.281 & 0.278 & 0.277 & 0.281 & 0.282 & 0.278 \\
 &  2 & 11 & 0.177 & 0.188 & 0.195 & 0.191 & 0.178 & 0.160 & 0.143 & 0.132 & 0.125 & 0.121 & 0.120 & 0.120 & 0.118 \\
 &  2 & 12 & 1.319 & 1.343 & 1.329 & 1.257 & 1.137 & 0.998 & 0.862 & 0.753 & 0.686 & 0.664 & 0.673 & 0.690 & 0.697 \\
 &  2 & 13 & 0.354 & 0.349 & 0.340 & 0.335 & 0.338 & 0.344 & 0.347 & 0.351 & 0.358 & 0.372 & 0.388 & 0.398 & 0.397 \\
 &  2 & 14 & 0.593 & 0.590 & 0.585 & 0.573 & 0.551 & 0.523 & 0.495 & 0.472 & 0.464 & 0.475 & 0.497 & 0.523 & 0.539 \\
 &  2 & 15 & 0.163 & 0.157 & 0.153 & 0.150 & 0.151 & 0.155 & 0.161 & 0.166 & 0.171 & 0.177 & 0.184 & 0.193 & 0.200 \\
 &  3 &  4 & 1.591 & 1.611 & 1.692 & 1.855 & 2.071 & 2.278 & 2.413 & 2.462 & 2.436 & 2.327 & 2.143 & 1.923 & 1.696 \\
 &  3 &  5 & 1.016 & 0.953 & 0.886 & 0.822 & 0.752 & 0.677 & 0.607 & 0.550 & 0.518 & 0.518 & 0.533 & 0.538 & 0.518 \\
 &  3 &  6 & 0.544 & 0.574 & 0.594 & 0.592 & 0.568 & 0.535 & 0.501 & 0.469 & 0.448 & 0.448 & 0.463 & 0.474 & 0.467 \\
 &  3 &  7 & 0.301 & 0.283 & 0.264 & 0.248 & 0.240 & 0.238 & 0.236 & 0.231 & 0.231 & 0.243 & 0.262 & 0.274 & 0.273 \\
 &  3 &  8 & 0.343 & 0.344 & 0.342 & 0.340 & 0.336 & 0.329 & 0.323 & 0.322 & 0.326 & 0.332 & 0.342 & 0.350 & 0.351 \\
 &  3 &  9 & 0.509 & 0.493 & 0.476 & 0.466 & 0.462 & 0.464 & 0.475 & 0.498 & 0.520 & 0.537 & 0.553 & 0.565 & 0.569 \\
 &  3 & 10 & 0.129 & 0.130 & 0.140 & 0.155 & 0.171 & 0.181 & 0.184 & 0.182 & 0.179 & 0.181 & 0.188 & 0.193 & 0.192 \\
 &  3 & 11 & 0.151 & 0.159 & 0.172 & 0.184 & 0.185 & 0.176 & 0.161 & 0.150 & 0.142 & 0.139 & 0.138 & 0.136 & 0.132 \\
 &  3 & 12 & 0.312 & 0.358 & 0.393 & 0.401 & 0.384 & 0.350 & 0.312 & 0.280 & 0.262 & 0.261 & 0.272 & 0.285 & 0.292 \\
 &  3 & 13 & 0.251 & 0.244 & 0.232 & 0.220 & 0.213 & 0.212 & 0.216 & 0.223 & 0.232 & 0.243 & 0.255 & 0.262 & 0.259 \\
 &  3 & 14 & 0.627 & 0.625 & 0.640 & 0.646 & 0.628 & 0.591 & 0.550 & 0.519 & 0.507 & 0.516 & 0.541 & 0.572 & 0.595 \\
 &  3 & 15 & 0.179 & 0.177 & 0.175 & 0.175 & 0.177 & 0.184 & 0.193 & 0.202 & 0.211 & 0.221 & 0.234 & 0.245 & 0.250 \\
 &  4 &  5 & 0.910 & 0.818 & 0.721 & 0.631 & 0.546 & 0.467 & 0.398 & 0.342 & 0.306 & 0.293 & 0.291 & 0.288 & 0.275 \\
 &  4 &  6 & 0.373 & 0.394 & 0.404 & 0.394 & 0.369 & 0.340 & 0.314 & 0.291 & 0.281 & 0.291 & 0.311 & 0.324 & 0.320 \\
 &  4 &  7 & 0.261 & 0.248 & 0.233 & 0.223 & 0.220 & 0.223 & 0.226 & 0.225 & 0.224 & 0.228 & 0.240 & 0.250 & 0.251 \\
 &  4 &  8 & 0.163 & 0.166 & 0.166 & 0.164 & 0.159 & 0.151 & 0.143 & 0.138 & 0.134 & 0.131 & 0.131 & 0.133 & 0.133 \\
 &  4 &  9 & 0.376 & 0.365 & 0.356 & 0.355 & 0.359 & 0.365 & 0.376 & 0.396 & 0.416 & 0.431 & 0.444 & 0.456 & 0.461 \\
 &  4 & 10 & 0.059 & 0.067 & 0.078 & 0.089 & 0.099 & 0.104 & 0.105 & 0.102 & 0.100 & 0.102 & 0.109 & 0.116 & 0.118 \\
 &  4 & 11 & 0.106 & 0.105 & 0.117 & 0.135 & 0.147 & 0.146 & 0.137 & 0.127 & 0.122 & 0.119 & 0.118 & 0.116 & 0.111 \\
 &  4 & 12 & 0.053 & 0.056 & 0.062 & 0.067 & 0.069 & 0.069 & 0.067 & 0.065 & 0.065 & 0.068 & 0.075 & 0.081 & 0.085 \\
 &  4 & 13 & 0.161 & 0.154 & 0.143 & 0.130 & 0.119 & 0.111 & 0.108 & 0.109 & 0.111 & 0.116 & 0.124 & 0.130 & 0.132 \\
 &  4 & 14 & 0.494 & 0.501 & 0.525 & 0.538 & 0.525 & 0.492 & 0.453 & 0.420 & 0.403 & 0.407 & 0.429 & 0.458 & 0.480 \\
 &  4 & 15 & 0.145 & 0.146 & 0.146 & 0.147 & 0.150 & 0.157 & 0.167 & 0.177 & 0.187 & 0.200 & 0.213 & 0.223 & 0.226 \\
 &  5 &  6 & 1.391 & 1.531 & 1.673 & 1.759 & 1.760 & 1.676 & 1.526 & 1.343 & 1.171 & 1.043 & 0.961 & 0.910 & 0.875 \\
 &  5 &  7 & 0.900 & 0.835 & 0.777 & 0.732 & 0.689 & 0.636 & 0.570 & 0.499 & 0.439 & 0.403 & 0.386 & 0.381 & 0.378 \\
 &  5 &  8 & 0.518 & 0.490 & 0.473 & 0.473 & 0.486 & 0.493 & 0.480 & 0.449 & 0.407 & 0.367 & 0.340 & 0.328 & 0.324 \\
 &  5 &  9 & 0.339 & 0.343 & 0.330 & 0.304 & 0.272 & 0.239 & 0.210 & 0.184 & 0.163 & 0.146 & 0.136 & 0.132 & 0.133 \\
 &  5 & 10 & 0.205 & 0.197 & 0.205 & 0.225 & 0.242 & 0.250 & 0.249 & 0.245 & 0.243 & 0.249 & 0.263 & 0.281 & 0.291 \\
 &  5 & 11 & 0.143 & 0.131 & 0.118 & 0.106 & 0.096 & 0.089 & 0.084 & 0.083 & 0.083 & 0.084 & 0.087 & 0.089 & 0.089 \\
 &  5 & 12 & 0.265 & 0.273 & 0.297 & 0.332 & 0.364 & 0.379 & 0.376 & 0.362 & 0.348 & 0.345 & 0.356 & 0.374 & 0.388 \\
 &  5 & 13 & 0.519 & 0.495 & 0.453 & 0.407 & 0.364 & 0.328 & 0.302 & 0.287 & 0.282 & 0.284 & 0.289 & 0.299 & 0.309 \\
 &  5 & 14 & 0.144 & 0.134 & 0.121 & 0.108 & 0.097 & 0.089 & 0.083 & 0.079 & 0.079 & 0.084 & 0.092 & 0.101 & 0.106 \\
 &  5 & 15 & 0.116 & 0.115 & 0.115 & 0.114 & 0.116 & 0.124 & 0.135 & 0.145 & 0.153 & 0.158 & 0.162 & 0.167 & 0.169 \\
 &  6 &  7 & 0.656 & 0.614 & 0.589 & 0.580 & 0.570 & 0.546 & 0.507 & 0.458 & 0.411 & 0.373 & 0.348 & 0.333 & 0.323 \\
 &  6 &  8 & 0.247 & 0.239 & 0.233 & 0.232 & 0.236 & 0.235 & 0.227 & 0.211 & 0.192 & 0.173 & 0.160 & 0.154 & 0.152 \\
 &  6 &  9 & 0.333 & 0.315 & 0.295 & 0.276 & 0.258 & 0.240 & 0.222 & 0.204 & 0.187 & 0.173 & 0.165 & 0.162 & 0.163 \\
 &  6 & 10 & 0.105 & 0.110 & 0.123 & 0.141 & 0.157 & 0.167 & 0.172 & 0.175 & 0.178 & 0.184 & 0.195 & 0.205 & 0.210 \\
 &  6 & 11 & 0.176 & 0.161 & 0.144 & 0.128 & 0.114 & 0.102 & 0.094 & 0.090 & 0.091 & 0.094 & 0.099 & 0.106 & 0.108 \\
 &  6 & 12 & 0.096 & 0.113 & 0.138 & 0.158 & 0.164 & 0.159 & 0.148 & 0.137 & 0.127 & 0.122 & 0.123 & 0.127 & 0.131 \\
 &  6 & 13 & 0.270 & 0.259 & 0.242 & 0.222 & 0.203 & 0.188 & 0.178 & 0.173 & 0.173 & 0.175 & 0.180 & 0.187 & 0.192 \\
 &  6 & 14 & 0.193 & 0.187 & 0.181 & 0.175 & 0.169 & 0.163 & 0.159 & 0.156 & 0.154 & 0.156 & 0.164 & 0.176 & 0.184 \\
 &  6 & 15 & 0.088 & 0.088 & 0.088 & 0.087 & 0.087 & 0.090 & 0.095 & 0.103 & 0.111 & 0.119 & 0.125 & 0.131 & 0.136 \\
 &  7 &  8 & 0.077 & 0.077 & 0.080 & 0.086 & 0.092 & 0.094 & 0.090 & 0.082 & 0.072 & 0.062 & 0.053 & 0.048 & 0.045 \\
 &  7 &  9 & 0.164 & 0.157 & 0.152 & 0.149 & 0.147 & 0.144 & 0.138 & 0.130 & 0.120 & 0.111 & 0.105 & 0.104 & 0.103 \\
 &  7 & 10 & 0.035 & 0.037 & 0.043 & 0.053 & 0.063 & 0.070 & 0.074 & 0.076 & 0.077 & 0.079 & 0.082 & 0.084 & 0.085 \\
 &  7 & 11 & 0.105 & 0.101 & 0.095 & 0.088 & 0.079 & 0.071 & 0.065 & 0.062 & 0.063 & 0.066 & 0.071 & 0.076 & 0.078 \\
 &  7 & 12 & 0.025 & 0.042 & 0.061 & 0.071 & 0.070 & 0.060 & 0.049 & 0.039 & 0.031 & 0.026 & 0.024 & 0.023 & 0.024 \\
 &  7 & 13 & 0.114 & 0.113 & 0.108 & 0.101 & 0.094 & 0.086 & 0.081 & 0.077 & 0.076 & 0.075 & 0.075 & 0.077 & 0.079 \\
 &  7 & 14 & 0.138 & 0.135 & 0.132 & 0.129 & 0.125 & 0.121 & 0.118 & 0.115 & 0.113 & 0.113 & 0.118 & 0.125 & 0.132 \\
 &  7 & 15 & 0.043 & 0.043 & 0.043 & 0.044 & 0.046 & 0.048 & 0.051 & 0.055 & 0.059 & 0.063 & 0.067 & 0.071 & 0.075 \\
 &  8 &  9 & 1.248 & 1.372 & 1.481 & 1.552 & 1.573 & 1.557 & 1.531 & 1.501 & 1.448 & 1.388 & 1.369 & 1.406 & 1.445 \\
 &  8 & 10 & 0.419 & 0.446 & 0.483 & 0.516 & 0.543 & 0.570 & 0.594 & 0.606 & 0.601 & 0.581 & 0.561 & 0.550 & 0.544 \\
 &  8 & 11 & 0.411 & 0.384 & 0.358 & 0.340 & 0.332 & 0.324 & 0.316 & 0.306 & 0.292 & 0.273 & 0.254 & 0.242 & 0.235 \\
 &  8 & 12 & 1.282 & 1.265 & 1.281 & 1.321 & 1.362 & 1.385 & 1.383 & 1.380 & 1.434 & 1.596 & 1.839 & 2.063 & 2.164 \\
 &  8 & 13 & 0.735 & 0.717 & 0.692 & 0.664 & 0.643 & 0.628 & 0.615 & 0.607 & 0.607 & 0.629 & 0.689 & 0.775 & 0.841 \\
 &  8 & 14 & 0.956 & 0.989 & 1.015 & 1.033 & 1.039 & 1.025 & 1.000 & 0.978 & 0.990 & 1.058 & 1.160 & 1.247 & 1.278 \\
 &  8 & 15 & 0.358 & 0.372 & 0.386 & 0.391 & 0.387 & 0.376 & 0.363 & 0.349 & 0.339 & 0.341 & 0.362 & 0.395 & 0.418 \\
 &  9 & 10 & 0.493 & 0.429 & 0.384 & 0.367 & 0.376 & 0.401 & 0.427 & 0.446 & 0.453 & 0.445 & 0.437 & 0.437 & 0.440 \\
 &  9 & 11 & 0.411 & 0.390 & 0.386 & 0.402 & 0.417 & 0.414 & 0.395 & 0.372 & 0.346 & 0.318 & 0.297 & 0.288 & 0.286 \\
 &  9 & 12 & 0.426 & 0.516 & 0.639 & 0.763 & 0.846 & 0.874 & 0.862 & 0.838 & 0.830 & 0.856 & 0.909 & 0.962 & 0.988 \\
 &  9 & 13 & 0.422 & 0.410 & 0.402 & 0.396 & 0.395 & 0.398 & 0.405 & 0.415 & 0.428 & 0.450 & 0.490 & 0.539 & 0.574 \\
 &  9 & 14 & 0.990 & 0.967 & 0.940 & 0.909 & 0.878 & 0.856 & 0.847 & 0.856 & 0.915 & 1.055 & 1.248 & 1.413 & 1.482 \\
 &  9 & 15 & 0.293 & 0.315 & 0.342 & 0.364 & 0.376 & 0.381 & 0.382 & 0.382 & 0.383 & 0.392 & 0.420 & 0.464 & 0.498 \\
 & 10 & 11 & 0.357 & 0.338 & 0.334 & 0.349 & 0.371 & 0.387 & 0.403 & 0.427 & 0.446 & 0.451 & 0.450 & 0.449 & 0.440 \\
 & 10 & 12 & 0.158 & 0.174 & 0.192 & 0.199 & 0.190 & 0.172 & 0.154 & 0.143 & 0.144 & 0.156 & 0.173 & 0.186 & 0.189 \\
 & 10 & 13 & 0.587 & 0.573 & 0.558 & 0.547 & 0.546 & 0.557 & 0.580 & 0.609 & 0.635 & 0.660 & 0.695 & 0.731 & 0.747 \\
 & 10 & 14 & 0.099 & 0.098 & 0.097 & 0.095 & 0.095 & 0.098 & 0.102 & 0.106 & 0.115 & 0.129 & 0.144 & 0.154 & 0.156 \\
 & 10 & 15 & 0.423 & 0.454 & 0.495 & 0.534 & 0.558 & 0.571 & 0.578 & 0.575 & 0.562 & 0.549 & 0.542 & 0.540 & 0.530 \\
 & 11 & 12 & 0.053 & 0.064 & 0.075 & 0.080 & 0.076 & 0.068 & 0.059 & 0.052 & 0.050 & 0.054 & 0.058 & 0.060 & 0.058 \\
 & 11 & 13 & 0.319 & 0.296 & 0.275 & 0.259 & 0.255 & 0.265 & 0.284 & 0.302 & 0.313 & 0.321 & 0.332 & 0.344 & 0.350 \\
 & 11 & 14 & 0.042 & 0.041 & 0.040 & 0.040 & 0.042 & 0.046 & 0.050 & 0.055 & 0.061 & 0.068 & 0.076 & 0.083 & 0.087 \\
 & 11 & 15 & 0.178 & 0.198 & 0.227 & 0.258 & 0.285 & 0.304 & 0.313 & 0.310 & 0.297 & 0.283 & 0.276 & 0.274 & 0.268 \\
 & 12 & 13 & 0.343 & 0.332 & 0.316 & 0.299 & 0.286 & 0.283 & 0.292 & 0.310 & 0.336 & 0.368 & 0.400 & 0.425 & 0.434 \\
 & 12 & 14 & 2.767 & 2.722 & 2.622 & 2.476 & 2.301 & 2.119 & 1.958 & 1.841 & 1.805 & 1.880 & 2.050 & 2.241 & 2.361 \\
 & 12 & 15 & 0.073 & 0.072 & 0.073 & 0.079 & 0.088 & 0.098 & 0.106 & 0.114 & 0.130 & 0.151 & 0.169 & 0.179 & 0.178 \\
 & 13 & 14 & 0.164 & 0.168 & 0.171 & 0.171 & 0.171 & 0.171 & 0.174 & 0.182 & 0.197 & 0.220 & 0.247 & 0.269 & 0.277 \\
 & 13 & 15 & 0.295 & 0.302 & 0.314 & 0.328 & 0.346 & 0.378 & 0.427 & 0.481 & 0.526 & 0.564 & 0.610 & 0.659 & 0.688 \\
 & 14 & 15 & 0.160 & 0.162 & 0.170 & 0.185 & 0.200 & 0.212 & 0.222 & 0.232 & 0.246 & 0.261 & 0.277 & 0.293 & 0.301 \\
\hline
\end{longtable}


\begin{thebibliography}{99}

\bibitem[\protect\citeauthoryear{Axelrod}{1980}]{AxelrodThesis1980}
Axelrod T.S., 1980, PhD thesis (UCRL-52994), University of California Santa Cruz

\bibitem[\protect\citeauthoryear{Badnell\ }{2011}]{AS2011}
Badnell N.R., 2011, Comput. Phys. Commun. 182, 1528

\bibitem[\protect\citeauthoryear{Baluteau \etal}{1995}]{BaluteauZMP1995}
Baluteau J.P., Zavagno A., Morisset C., P\'{e}quignot D., 1995, A\&A, 303, 175

\bibitem[\protect\citeauthoryear{Berrington \etal}{1974}]{BerringtonBCCRT1974}
Berrington K.A., Burke P.G., Chang J.J., Chivers A.T., Robb W.D., Taylor K.T., 1974, Comput. Phys.
Commun., 8, 149

\bibitem[\protect\citeauthoryear{Berrington \etal}{1987}]{BerringtonBBSSTY1987}
Berrington K.A., Burke P.G., Butler K., Seaton M.J., Storey P.J., Taylor K.T., Yu Yan., 1987, J.
Phys. B, 20, 6379

\bibitem[\protect\citeauthoryear{Berrington \etal}{1995}]{BerringtonEN1995}
Berrington K.A., Eissner W.B., Norrington P.H., 1995, Comput. Phys. Commun., 92, 290

\bibitem[\protect\citeauthoryear{Bowers \etal}{1997}]{BowersMGWPe1997}
Bowers E.J.C., Meikle W.P.S., Geballe T.R., \etal, 1997,  MNRAS, 290, 663

\bibitem[\protect\citeauthoryear{Childress \etal}{2015}]{ChildressHSSMe2015}
Childress M.J., Hillier D.J., Seitenzahl I., \etal, 2015, MNRAS, 454, 3816

\bibitem[\protect\citeauthoryear{Churazov \etal}{2014}]{ChurazovSIKJe2014}
Churazov E., Sunyaev R., Isern J., \etal, 2014, Nature, 512, 406

\bibitem[\protect\citeauthoryear{Colgate \& McKee}{1969}]{ColgateM1969}
Colgate S.A., McKee C., 1969, ApJ, 157, 623

\bibitem[\protect\citeauthoryear{Dessart \etal}{2014}]{DessartHBK2014}
Dessart L., Hillier D.J., Blondin S., Khokhlov A., 2014, MNRAS, 439, 3114

\bibitem[\protect\citeauthoryear{Eissner \etal}{1974}]{EissnerJN1974}
Eissner W., Jones M., Nussbaumer H., 1974, Comput. Phys. Commun., 8, 270

\bibitem[\protect\citeauthoryear{Fang \& Liu}{2011}]{FangL2011}
Fang X., Liu X.-W., 2011, MNRAS, 415, 181

\bibitem[\protect\citeauthoryear{Fivet \etal}{2016}]{FivetQB2016}
Fivet V., Quinet P., Bautista M.A., 2016, A\&A, 585, A121

\bibitem[\protect\citeauthoryear{Hansen \etal}{1984}]{HansenRU1984}
Hansen J.E., Raassen A.J.J., Uylings P.H.M., 1984, ApJ, 277, 435

\bibitem[\protect\citeauthoryear{Hummer \etal}{1993}]{HummerBEPST1993}
Hummer D.G., Berrington K.A., Eissner W., Pradhan A.K., Saraph H.E., Tully J.A., 1993, A\&A, 279,
298

\bibitem[\protect\citeauthoryear{Kuchner \etal}{1994}]{KuchnerKPL1994}
Kuchner M.J., Kirshner R.P., Pinto P.A., Leibundgut B., 1994, ApJ, 426, L89

\bibitem[\protect\citeauthoryear{Liu \etal}{1997}]{LiuJSQSP1997}
Liu W., Jeffery D.J., Schultz D.R., Quinet P., Shaw J., Pindzola M.S., 1997, ApJ, 489, L141

\bibitem[\protect\citeauthoryear{Nussbaumer \& Storey\ }{1978}]{NussbaumerS1978}
Nussbaumer H., Storey P.J., 1978, A\&A, 64, 139

\bibitem[\protect\citeauthoryear{Nussbaumer \& Storey}{1988}]{NussbaumerS1988}
Nussbaumer H., Storey P.J., 1988, A\&A, 200, L25

\bibitem[\protect\citeauthoryear{Pottasch \& Surendiranath}{2005}]{PottaschS2005b}
Pottasch S.R., Surendiranath R., 2005, A\&A, 444, 861

\bibitem[\protect\citeauthoryear{Storey \& Hummer}{1995}]{StoreyH95}
Storey P.J., Hummer D.G., 1995, MNRAS, 272, 41

\bibitem[\protect\citeauthoryear{Storey \etal}{2016}]{StoreyZS15}
Storey P.J., Zeippen C.J., Sochi T., 2016, MNRAS, 456, 1974

\bibitem[\protect\citeauthoryear{Sugar \& Corliss}{1981}]{SugarC1981}
Sugar J., Corliss C., 1981,  J. Phys. Chem. Ref. Data, 10, 1097

\bibitem[\protect\citeauthoryear{Sugar \& Corliss}{1985}]{SugarC1985}
Sugar J., Corliss C., 1985,  J. Phys. Chem. Ref. Data, 14, 1

\bibitem[\protect\citeauthoryear{Sunderland \etal}{2002}]{SunderlandNBB2002}
Sunderland A.G., Noble C.J., Burke V.M., Burke P.G., 2002, Comput. Phys. Commun., 145, 311

\bibitem[\protect\citeauthoryear{Tankosi\'{c} \etal}{2003}]{TankosicPD2003}
Tankosi\'{c} D., Popovi\'{c} L.C., Dimitrijevi\'{c} M.S., 2003, A\&A, 399, 795

\bibitem[\protect\citeauthoryear{Varani \etal}{1990}]{VaraniMSA1990}
Varani G.F., Meikle W.P.S., Spyromilio J., Allen D.A., 1990, MNRAS, 245, 570

\bibitem[\protect\citeauthoryear{Wang \& Liu}{2007}]{WangL2007}
Wang W., Liu X.-W., 2007, MNRAS, 381, 669

\bibitem[\protect\citeauthoryear{Zhang \etal}{2005}]{ZhangLLPB2005}
Zhang Y., Liu X.-W., Luo S.-G., P\'{e}quignot D., Barlow M.J., 2005, A\&A, 442, 249

\end{thebibliography}
\end{document}